\documentclass[a4paper,fleqn,usenatbib]{mnras}
\usepackage{newtxtext,newtxmath}
\usepackage[T1]{fontenc}
\usepackage{ae,aecompl}
\usepackage{graphics}
\usepackage{graphicx}
\usepackage{amsmath}	
\usepackage{booktabs,caption}

\title[4U 1626-67]{The ongoing spin-down episode of 4U 1626-67}

\author[Tobrej et al.]{
Mohammed Tobrej,$^{1}$\thanks{tabrez.md565@gmail.com}
Ruchi Tamang,$^{1}$\thanks{ruchitamang76@gmail.com}
Binay Rai,$^{1}$\thanks{binayrai21@gmail.com}
Manoj Ghising,$^{1}$\thanks{manojghising26@gmail.com}
\newauthor
Bikash Chandra Paul$^{1}$\thanks{bcpaul@associates.iucaa.in}
\\
$^{1}$Department of Physics, North Bengal University, Siliguri, Darjeeling, WB, 734013, India
\\
}

\pubyear{2023}

\begin{document}
\label{firstpage}
\pagerange{\pageref{firstpage}--\pageref{lastpage}}
\maketitle

\begin{abstract}
We report the X-ray characteristics of the persistent X-ray pulsar 4U 1626-67 using simultaneous NuSTAR and NICER observations. The X-ray pulsar 4U 1626-67 has just encountered a torque reversal in 2023 and is presently in the spin-down state. We have examined the temporal and spectral characteristics of the source during its ongoing spin-down episode. The pulse profiles of the source are characterized by multiple substructures at lower energies and a wide asymmetric single-peaked structure at higher energies. The pulse fraction follows an overall increasing trend with energy. We confirm the existence of mHz quasi-periodic oscillation (QPO) exclusively during the current spin-down phase in all the observations. The source is spinning down at  0.00045(4)\; s\; $yr^{-1}$. The broadband spectrum during this phase is described by empirical NPEX model and a soft blackbody component with kT $\sim$ 0.25 keV. In addition to the iron emission line, we also confirm the presence of cyclotron line at $\sim$ 36 keV. The source flux continues to decrease during the current spin-down phase, and the corresponding luminosity $\sim$ (3.3-4.9)\;$\times 10^{36}\; ergs\; s^{-1}$ lies in the intermediate range of accreting X-ray pulsars that may be associated with a hybrid accretion geometry. The magnetic field strengths estimated using the cyclotron line measurements and QPO frequency are consistent. The evolution of the spectral parameters relative to the pulsed phase is examined using phase-resolved spectroscopy.
\end{abstract}


%
%

\section{Introduction}
The persistent source 4U 1626-67 is an accretion-powered X-ray pulsar in an ultra-compact binary characterized by an orbital period of 42 minutes \citep{Middleditch, Chakrabarty}. It was discovered by Uhuru \citep{Giacconi} and was found to spin with a period $\sim$ 7.68 s \citep{Rappaport}. It maintained a constant spin period until $\sim$ 1990 \citep{Chakrabarty}, when it started a spin-down phase. During this phase, the source luminosity was found to diminish. The pulsar started spinning up in 2008 after experiencing a torque reversal \citep{Camero-Arranz, Camero-Arranzz}, accompanied by a significant increase in flux, restoring the source to its luminosity to that before 1990. Further, the source encountered another torque-reversal that was associated with an increase in the source luminosity \citep{Jain}. In the most recent instance, 4U 1626-67 encountered another torque reversal to a spin-down phase \citep{Jenke} which was further supported by \cite{Sharma}.

The torque state and energy significantly impact the pulse profile of 4U 1626-67 \citep{Beri}. The power density spectrum (PDS) of the source is known to be associated with torque-dependent mHz quasi-periodic oscillations (QPOs) \citep{Kaur, Jain, Beri, Shinoda, Owens, Kommers, Chakrabarty, Sharma}. In X-ray binaries, QPOs are typically known to be connected to the spinning of the inner accretion disk \citep{Paul}. QPOs in the power spectrum may be caused by any unevenly distributed materials or material blobs in the inner disk. This provides essential details about the interaction of accretion disk and neutron star magnetosphere in the case of accretion-powered X-ray pulsars. QPOs are observed at frequencies ranging from a few Hz to a few hundred Hz in black hole X-ray binaries and low magnetic field neutron stars. Neutron star systems with high magnetic fields are associated with only low-frequency QPOs, ranging in the limit 10 mHz to about 1 Hz. 

The X-ray spectra of 4U 1626-67 revealed complex emission features in the soft energy band. At $\sim$ 1 keV, an emission line from Ne X was observed, along with an emission from O VIII at $\sim$0.6 keV \citep{Angelini}. During the spin-down phase, high-resolution spectroscopy carried out by Chandra observations revealed that the hydrogenic Ne X and O VIII features are double-peaked in nature, indicating an accretion disk origin, and also verified emission lines from the He-like lines of Ne IX and O VII \citep{Schulz, Krauss}. The equivalent width and intensity of Ne and O emission lines vary significantly during the spin-up phase \citep{Camero-Arranzz}. Also, the pulse phase dependence of these emission lines suggests a warped structure in the accretion disk \citep{Beri}. The presence of the Fe-K $\alpha$ emission line, which has only been observed during the spin-up phase, is known to be strongly correlated with the luminosity \citep{Koliopanos}. 

The X-ray continuum of 4U 1626-67 is known to be associated with distinct absorption features. A narrow 37 keV cyclotron resonance scattering feature (CRSF) was first detected by \cite {Orlandini} which was further confirmed by \cite{Coburn, Iwakiri} and also reported the detection of its harmonic. During the second spin-up phase of the source, \cite{Iwakirii, D'Ai} reported the asymmetry in the cyclotron line profile. CRSFs are characteristic features  known to arise due to resonant scattering of hard X-ray photons and electrons in quantized energy levels \citep{Meszaros}. The separation between the energy levels represents the centroid energy of the cyclotron. The cyclotron line energy and the magnetic field strength are related as 

 $E_{cyc}$ = 11.6 $\times\; B_{12}$ (keV), where $ B_{12}$ is the magnetic field in units of $10^{12}$ G.
 
In this paper, we report a detailed study of the temporal and spectral features of the accretion-powered pulsar 4U 1626-67 using simultaneous NuSTAR and NICER observations. We present the observations and data reduction in Section 2, Timing analysis in Section 3, Spectral Analysis in Section 4, followed by the discussions of the results in Section 5. 
\section{Observation and Data reduction}
Various observatories recently observed the source 4U 1626-67. We have used concurrent data (Refer Table \ref{1}) from the Nuclear Spectroscopic Telescope Array (NuSTAR), Neutron Star Interior Composition Explorer (NICER), and some relevant observations from Fermi/GBM for our work. The simultaneous NuSTAR and NICER observations are now referred as Observation I to Observation IV respectively.

\subsection{NuSTAR}
The first hard X-ray focusing telescope, known as NuSTAR, is sensitive in the broad energy range of (3-79 keV). It is composed of two separate, co-aligned, grazing incident telescopes that are apparently similar but unidentical. Each telescope is designated by its focal plane module (FPMA or FPMB), which is made up of a solid state CdZnTe detector that has been pixelated \citep{Harrison}. NuSTAR is well suited for examining the distinctive features in X-ray binary systems since it has exceptional spectral resolution and sensitivity.   The mission-specific NUPIPELINE was used to build the clean event files that were used for the analysis. The obtained clean event files were read using the XSELECT tool. DS9 \footnote{\url{https://sites.google.com/cfa.harvard.edu/saoimageds9}}, an astronomical imaging and data visualization tool, was used to observe the image. Spectra and light curves were  extracted using a circular region of $80^{\prime \prime}$ as the source region and backgrounds were generated from an $80^{\prime \prime}$ circular region taken away from the source. The mission-specific task NUPRODUCTS was utilized to obtain the required light curve and spectra from the source and background files. Using FTOOL LCMATH, the background correction for the light curve was completed. The data were barycentered to the solar system frame by the FTOOL BARYCORR. The spectral fitting was performed in XSPEC version 12.13.0 \citep{Arnaud}. The data reduction and analysis is performed using HEASOFT  v6.31 \footnote{\url{https://heasarc.gsfc.nasa.gov/docs/software/heasoft/download.html}} along with NuSTAR CALDB v. 20230124.

\subsection{NICER}
To explore neutron star systems using soft Xray timing, the International Space Station (IIS) external payload NICER was developed \citep{29}. It consists of 56 aligned Focal Plane Modules (FPMs), 52 of which are currently working.  The two detectors (14, 34) are linked with an increased detector noise. A thermoelectric cooler, preamplifier, and detector make up each FPM. We have employed an X-ray Timing Instrument (XTI) that is sensitive in the (0.2-12.0) keV energy range. Using NICERDAS v10 and CALDBv XTI20221001, we completed the routine data screening and reduction of NICER observations and the tool nibackgen3C50 v7 \citep{30} was used for the background estimation corresponding to each observation. NICERL2 was utilized to filter the clean event files. In order to extract the necessary source light curves, PHA, response, and ARF files, the filtered event files were loaded into XSELECT. The spectra in the soft energy range (0.7-10) keV is considered for fitting. In order to minimize the effect of low-energy noises, we ignored the spectra below 0.7 keV. Due to background contamination, we excluded the spectra above 10 keV.

\subsection{Fermi/GBM}
Launched in 2008, the Fermi Gamma Ray Space Telescope investigates gamma-ray sources. It has a Gamma-ray Burst Monitor (GBM) that is effective between 20 MeV and 300 GeV \citep{Meegan}. We used the archival data of the source for our analysis. The spin frequency derivative was obtained using the spin frequency supplied by the FERMI GBM Accreting Pulsars Program (GAPP). The slope of the plot, together with the error associated with it, indicate the rate at which spin frequency changes, with  a 1 $\sigma$ measurement uncertainty.


\begin{table*}
\begin{center}
\begin{tabular}{cccccc}
\hline
Observation Date & Observation ID (NuSTAR) & Exposure (s)& Observation ID (NICER) & Exposure (s)    &	 
     \\	
\hline
2023-05-02 &90901318002 & 27422& 6203800107 &		1391	 \\
2023-06-04 &90901318006 &18468& 6203800130 & 		1210   \\
2023-06-22 &90901318008 &22332 & 6203800142 &      4845    \\
2023-07-05 &90901318010 &18449 & 6203800149 &     1947    \\

\hline
\end{tabular}

\caption{NuSTAR and NICER observations with date of observation, observation IDs, and exposure.}  
\label{1}
\end{center}

\end{table*}

\section{Timing Analysis}
We consider NuSTAR light curves with a binning of 0.01 s for performing the temporal analysis of the source. The required light curves for observation are generated using FTOOL LCURVE. The Power density spectra (PDS) of 4U 1626-67 corresponding to all the observations are obtained using the POWSPEC tool of XRONOS sub-package of ftools \citep{Blackburn}. Sharp peaks at $\sim$ 0.13 Hz and its harmonics corresponding to a spin period of $\sim$7.7s are observed in all the observations. In addition, a broad peak denoting a quasi-periodic oscillation (QPO) is observed in the PDS corresponding to all the observations. 

Generally, QPOs may be described by a Lorentzian,

 $P_{\nu}$ $\propto\frac{\lambda}{[(\nu-\nu_{o})^{2}+(\frac{\lambda}{2})^{2}]}$, where $\nu_{o}$ and $\lambda$ denote the centroid frequency and full width at half maximum (FWHM) respectively. The ratio between the QPO centroid frequency and FWHM gives a measure for the coherence/sharpness of the QPO and is known as the quality factor \citep{Belloni, van der Klis}. The strength (or variance) of a signal, is directly proportional to the integrated power (P) of its contribution to the power spectrum. This is expressed in terms of the signal's fractional root-mean-squared (rms) amplitude (r $\propto P^{1/2}$), which represents the signal's amplitude as a percentage of the total source flux. The QPO signal observed in the present study is characterized by an estimated average centroid frequency of $\sim$ 46.85 mHz, r.m.s. amplitude $\sim$ (16.54 $\pm$1.53) $\%$ and quality factor  $\sim$ 5.97  which is consistent with the recent reports of \cite{Sharma} (Refer Table \ref{2}). The PDS of the source corresponding to observation II is presented in Figure \ref{1}. The pulse period of 4U 1626-67 was roughly estimated using the Fast Fourier Transform (FFT) of the light curve. The epoch-folding technique \citep{16,17} based on $\chi^{2}$ maximization was implemented for determining the spin period of the source precisely. The best period of the source corresponding to each observation was estimated to be 7.668037(7), 7.668051(4), 7.668032(5), and 7.668048(7), respectively. The uncertainty in the pulsation measurements was estimated by the method given by \cite{Boldin} by generating 1000 simulated light curves. The pulsation for each curve was estimated, and the standard deviation and errors were computed for approximating the uncertainty in the measurement. The best pulse period of the source is estimated using the tool EFSEARCH and the folded pulse profile of the source is obtained using the tool EFOLD. Similarly, we have considered 0.018 s binned NICER light curves for the required temporal study of the source. We have also used the Fermi/Gamma-ray Burst Monitor (GBM) data \citep{Malacariaa} for studying the spin frequency history of the source alongside the estimated values based on our observations. It is observed that the spin frequency of 4U 1626-67 encountered a torque reversal approximately at MJD 60020 (March 17, 2023) \citep{Jenke}. The source is currently spinning down at the rate 0.00045(4) s\;$yr^{-1}$. The spin period evolution of the source is presented in Figure \ref{2}.
\begin{table}
\begin{center}
\begin{tabular}{cccccccccc}
\hline
QPO Parameters	&	Obs I	&	Obs II	&	Obs III	&	Obs IV	\\
\hline									
Position (mHz)	&46.82$\pm$0.84		&46.78$\pm$0.81		&46.93$\pm$0.78		&46.85$\pm$0.91	\\
R.m.s amplitude ($\%$)	&16.12$\pm$1.52		&16.44$\pm$1.45		&16.75$\pm$1.56		&16.87$\pm$1.59		\\
Quality factor	&	5.93	& 5.91		&	6.04	&	5.99	\\

\hline
\end{tabular}

\caption{Details of QPO parameters (position, r.m.s amplitude, and quality factor) corresponding to the four NuSTAR observations.}  
\label{2}
\end{center}

\end{table}
\begin{figure}

\begin{center}
\includegraphics[angle=270,scale=0.3]{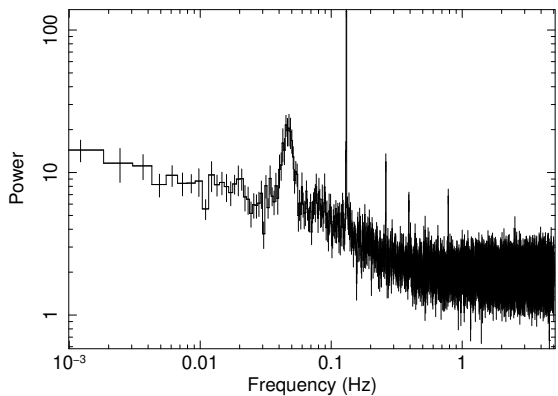}
\end{center}

\caption{Power Density Spectrum of 4U 1626-67 in (3-79) keV energy range corresponding to observation II using NuSTAR. The broad peak corrresponds the quasi-periodic signal.}
\label{1}
\end{figure}

\begin{figure}

\begin{center}
\includegraphics[angle=0,scale=0.29]{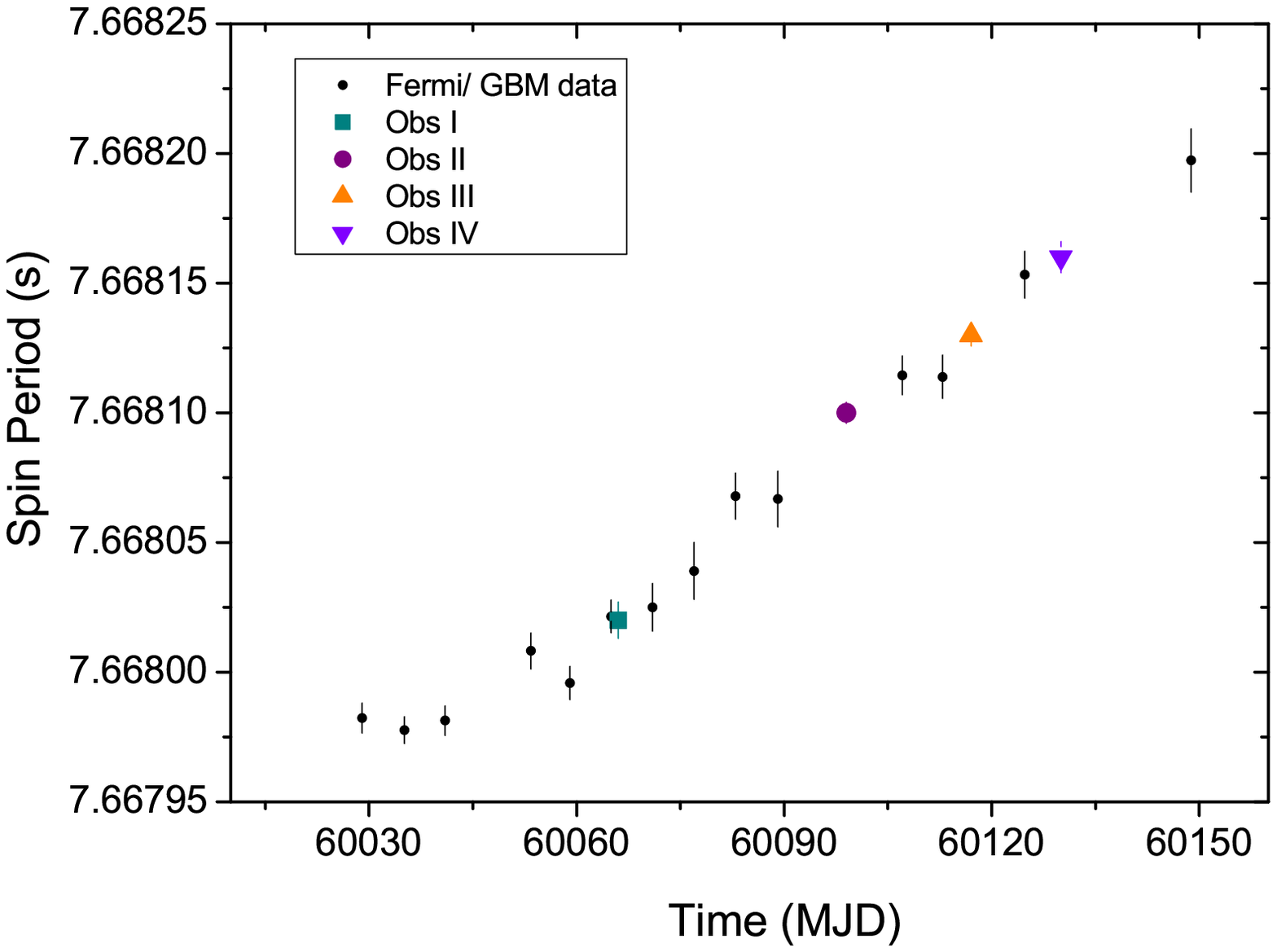}
\end{center}
\caption{Spin period evolution of 4U 1626-67 from Fermi-GBM along with NuSTAR observations from 2023-03-26 to 2023-07-23 during the current spin-down episode.}

\label{2}
\end{figure}
 
\subsection{Light curves, Pulse profiles and Pulse fraction}
\begin{figure}

\begin{center}
\includegraphics[angle=0,scale=0.3]{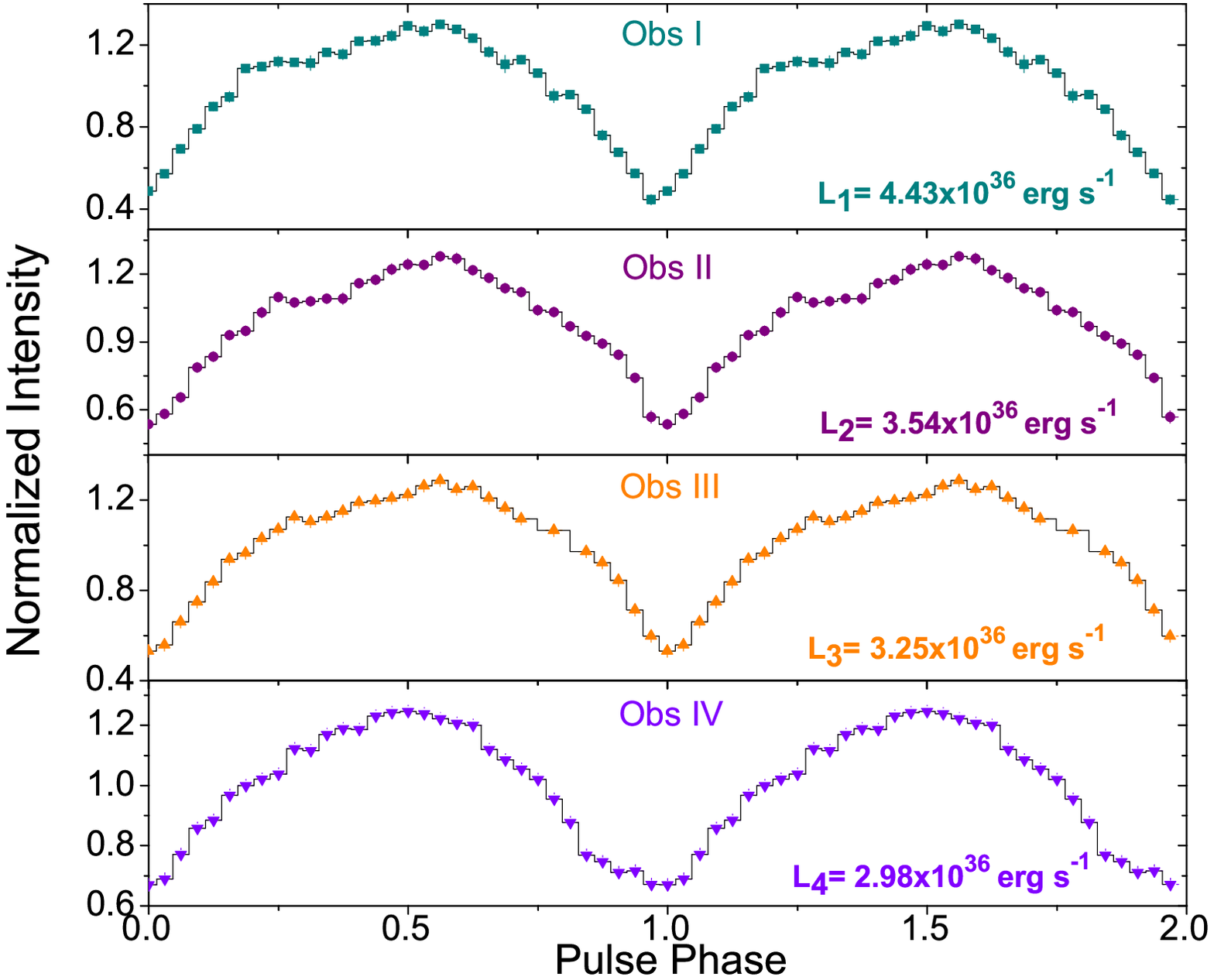}
\end{center}

\caption{Folded pulse profiles of 4U 1626-67 in (3-79) keV energy range using 32 bins. The pulse profiles corresponding to the four NuSTAR observations at different luminosities are normalized about the average count rate.}
\label{3}
\end{figure}
\begin{figure}

\begin{center}
\includegraphics[angle=0,scale=0.3]{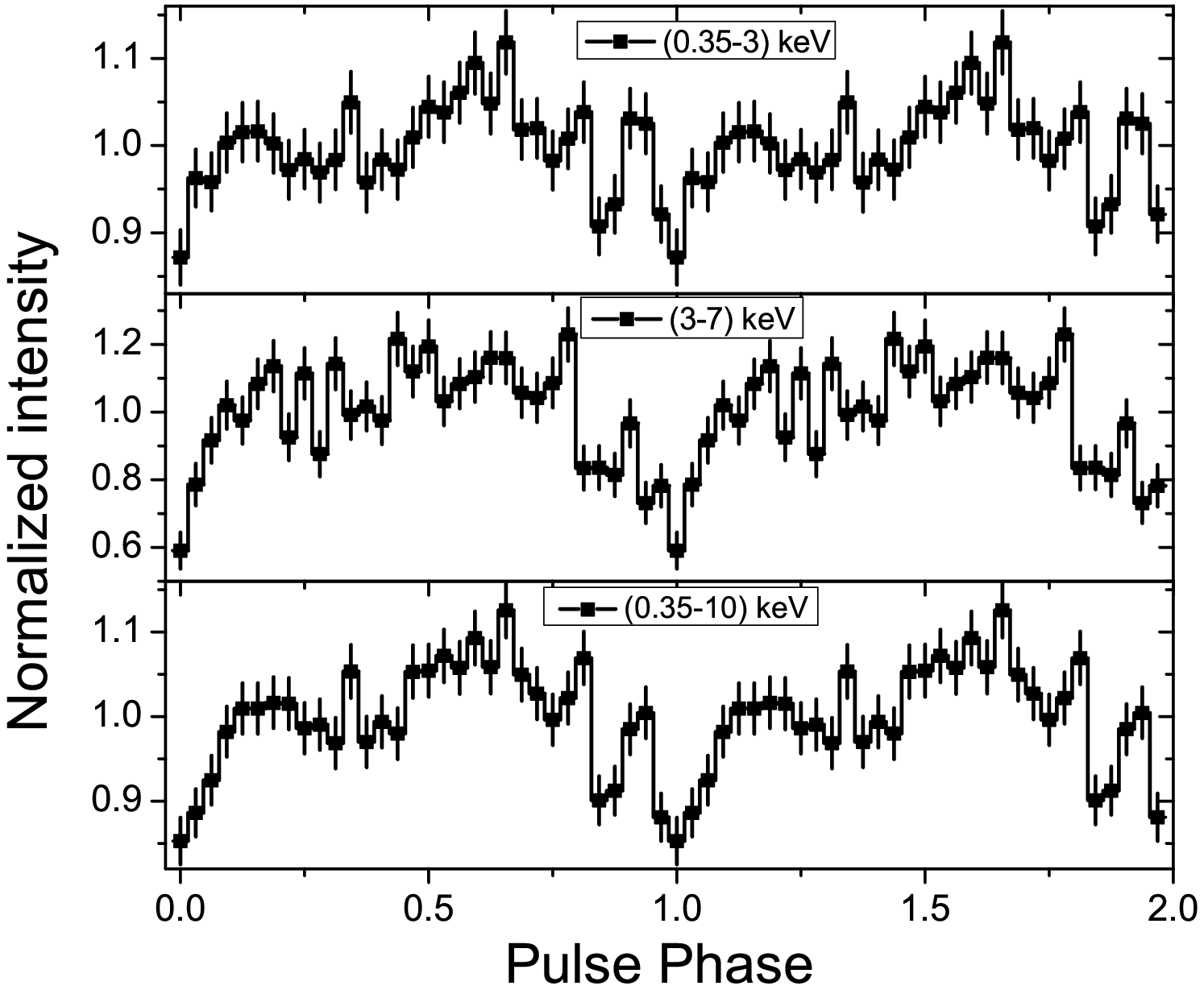}
\end{center}

\caption{Folded pulse profiles of 4U 1626-67 for NICER observation (Obs-I) at different energies in the soft energy band.}
\label{4}
\end{figure}
\begin{figure*}
\begin{minipage}{0.3\textwidth}

\includegraphics[angle=0,scale=0.25]{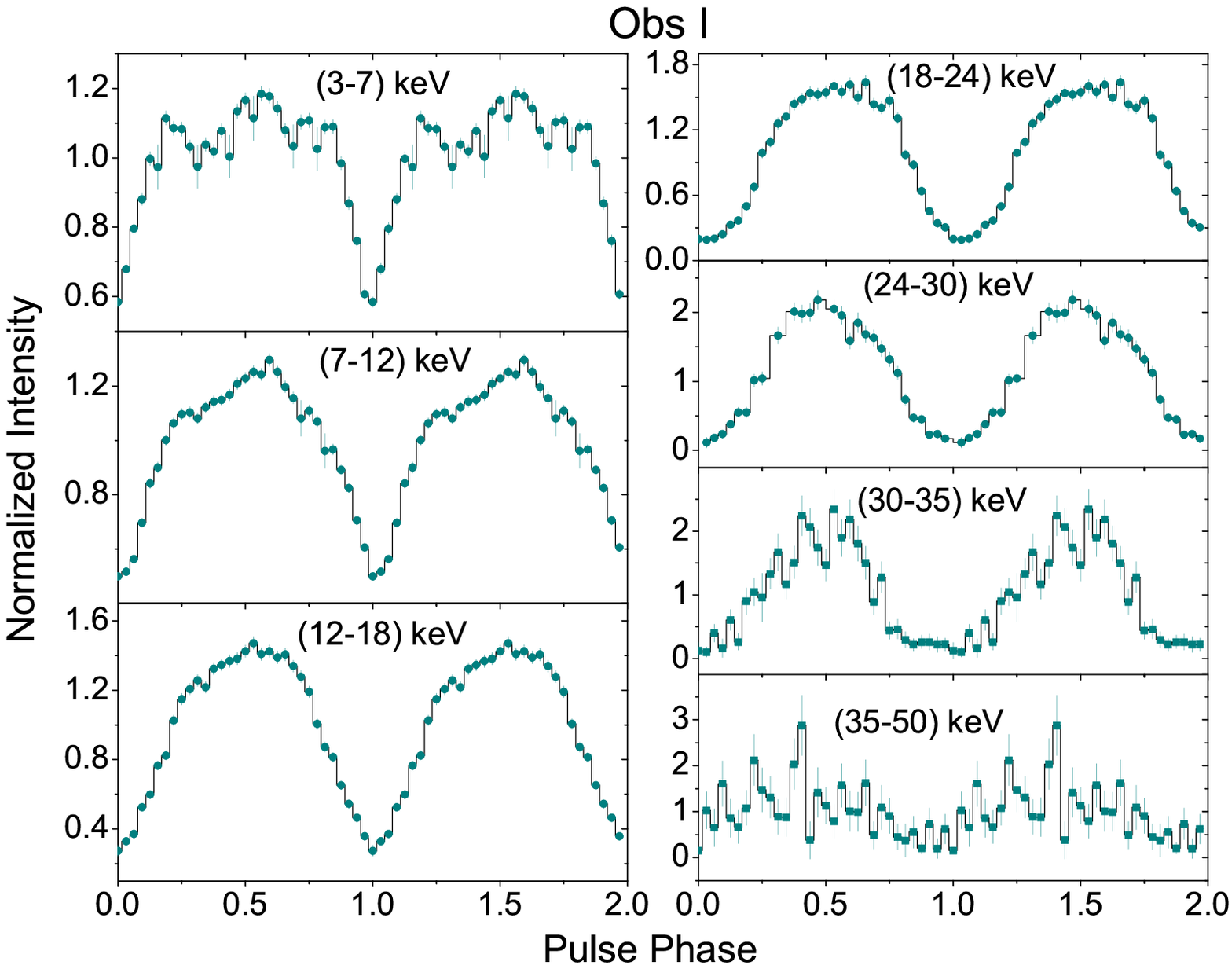}
\end{minipage}
\hspace{0.15\linewidth}
\begin{minipage}{0.3\textwidth}
\includegraphics[angle=0,scale=0.25]{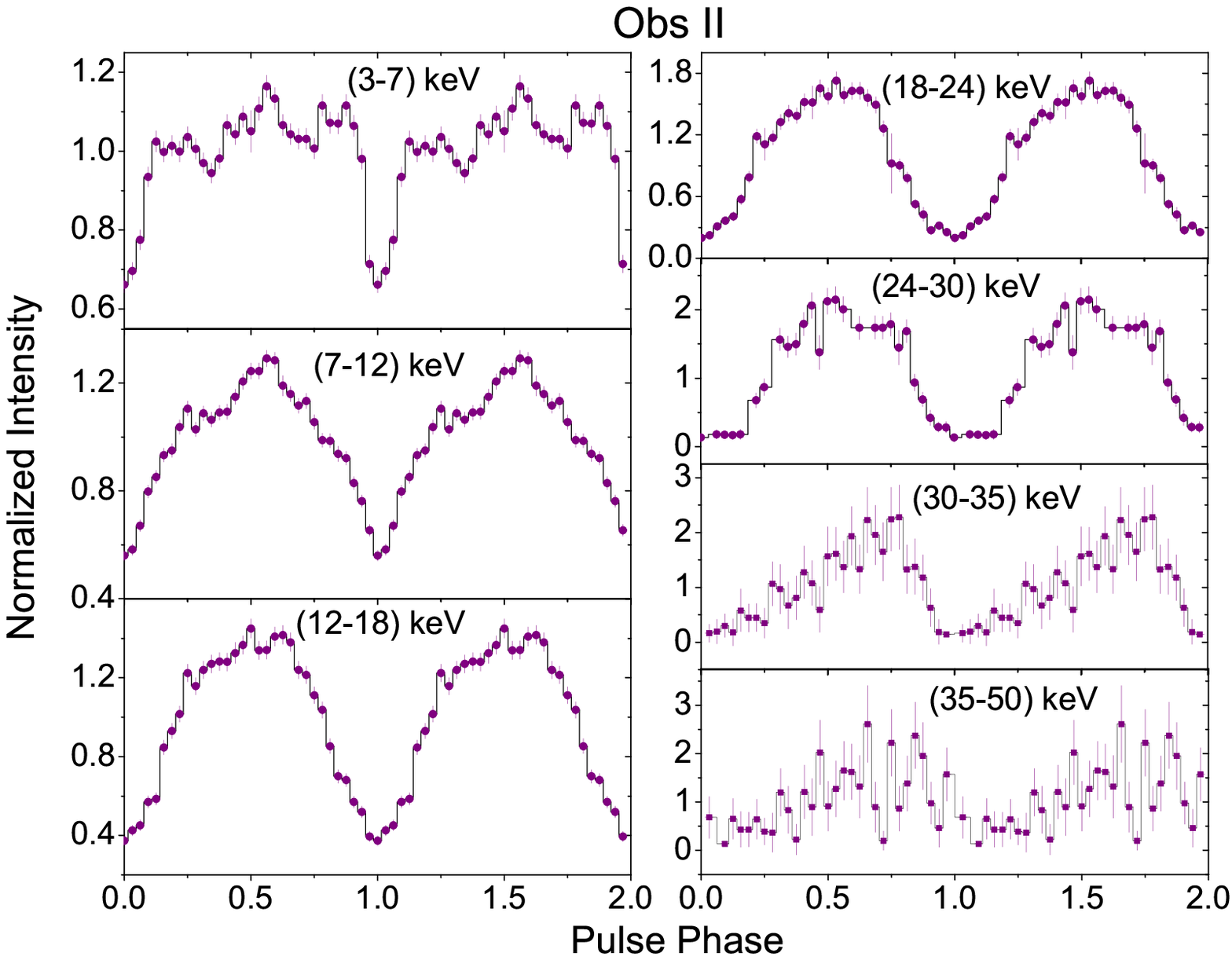}
\end{minipage}
\hspace{0.15\linewidth}
\begin{minipage}{0.3\textwidth}
\includegraphics[angle=0,scale=0.25]{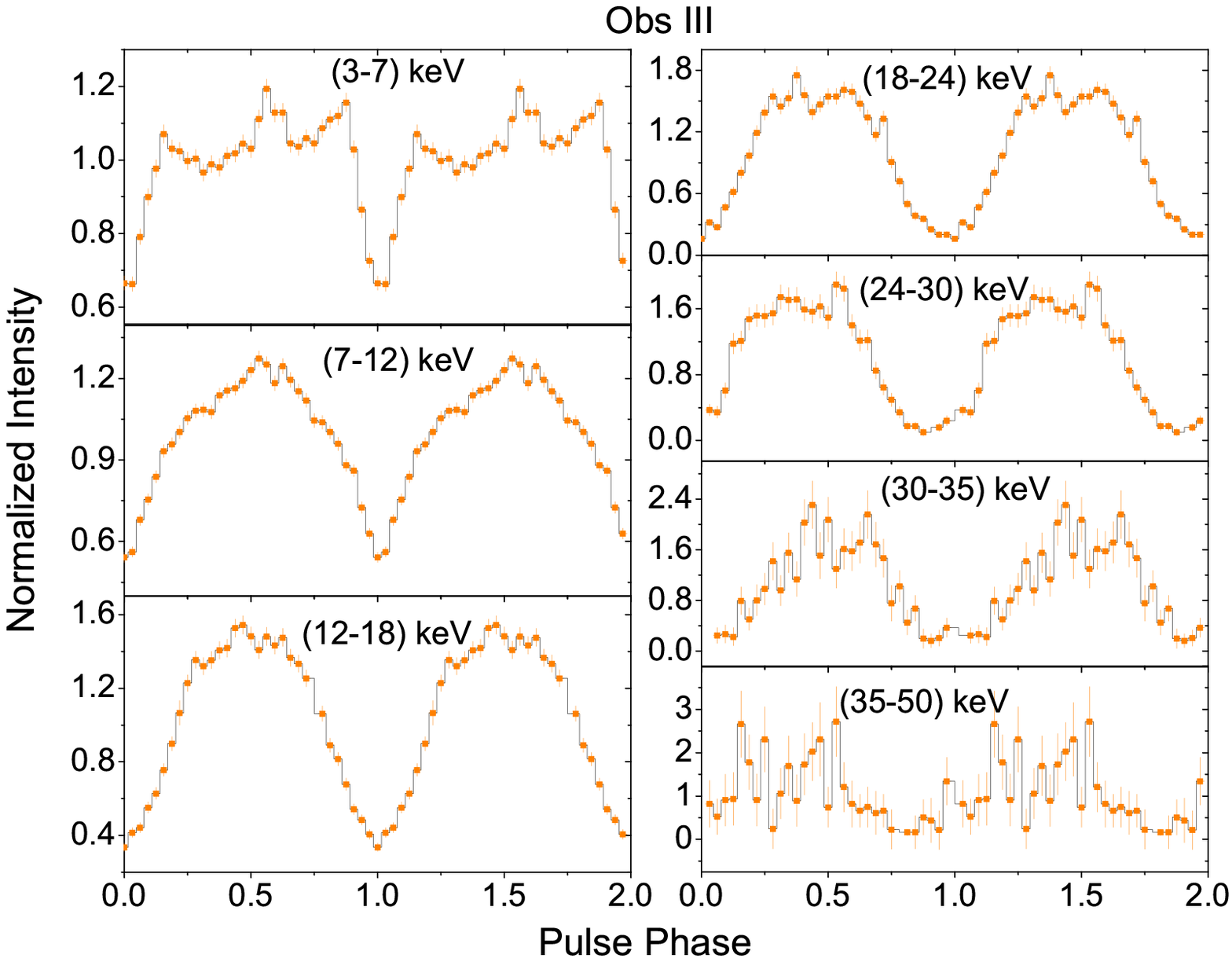}
\end{minipage}
\hspace{0.15\linewidth}
\begin{minipage}{0.3\textwidth}
\includegraphics[angle=0,scale=0.25]{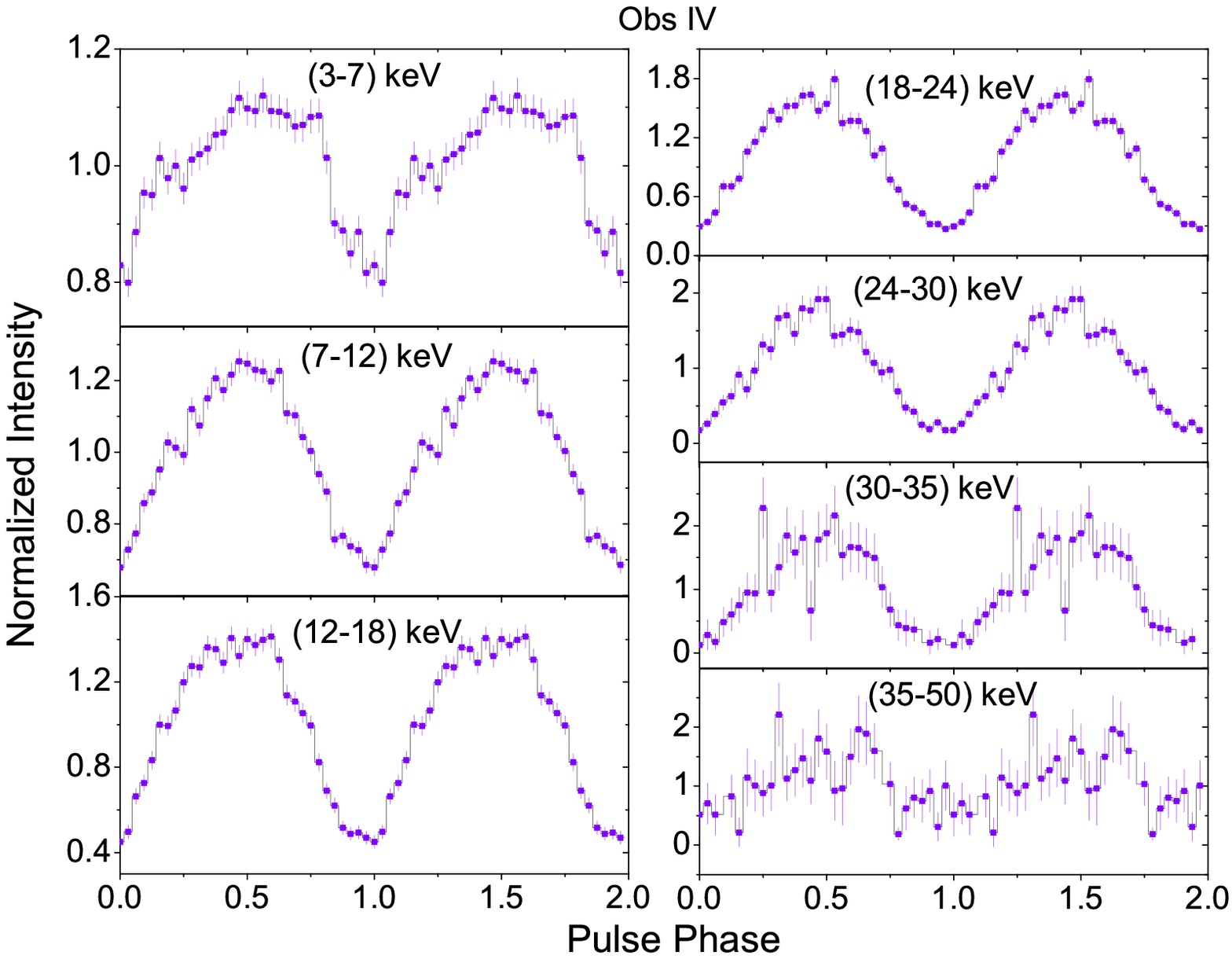}
\end{minipage}
\hspace{0.15\linewidth}
\caption{NuSTAR energy-resolved pulse profiles corresponding to observations I, II, III, \&IV respectively.}
\label{5}
\end{figure*}
\begin{figure}

\begin{center}
\includegraphics[angle=0,scale=0.3]{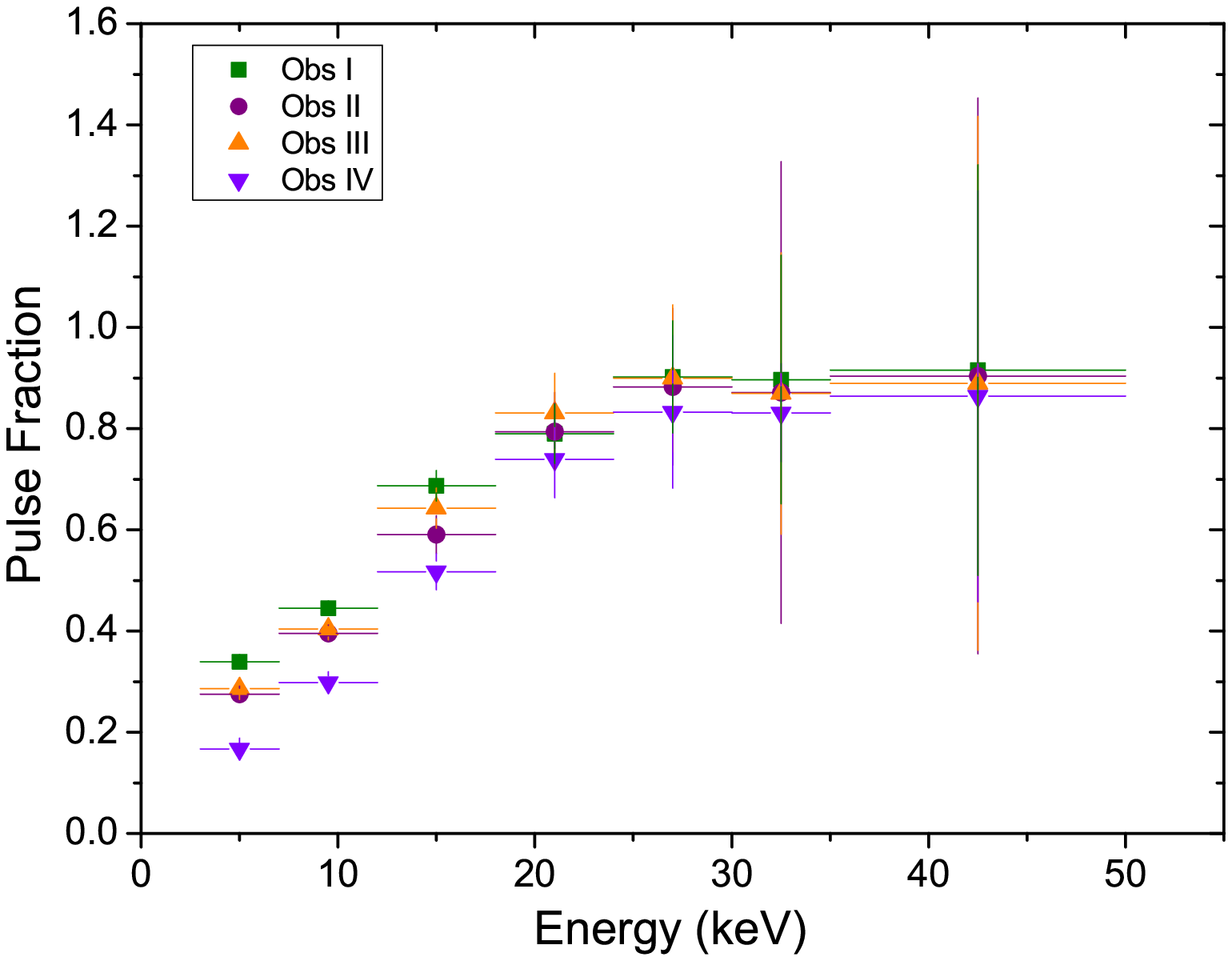}
\end{center}
\caption{Variation of pulse fraction with energy using NuSTAR observations.}
\label{6}
\end{figure}

The simultaneous NuSTAR and NICER observations are utilized to explore the X-ray characteristics of the source. The emission pattern at various energies are studied by generating the energy-resolved pulse profiles. The pulse profiles of the source corresponding to the four NuSTAR observations in (3-79) keV energy range do not exhibit much variability and are found to peak around the phase bin $\sim 0.5$. The energy-resolved pulse profiles are generated by resolving the light curve in the (3-79) keV energy range into several energy bands. The pulse profiles are found to evolve with energy. The pulse profiles are folded by a specific time zero-point $T_0$ (folding epoch), so the minimum flux bin is at phase point 0.0. At lower energies, multiple sub-structures are observed in the pulse profiles. At higher energies, the narrow sub-structures observed at low energies broaden, and the shape is primarily a broad asymmetric peak. With increasing energy, the pulse profiles exhibit a relative increase in the amplitude of the peak emission. As NuSTAR is not sensitive at lower energies, we have generated the pulse profiles in the soft energy range using NICER. It isn't easy  to assess the extent of modulation across the different energy bands due to variation in the count rates. The evolution of the pulse profiles infer an altering activity in the accretion geometry of the pulsar. The folded pulse profiles of the
source corresponding to each NuSTAR observation in (3-79) keV energy range is presented in Figure \ref{3}. The energy resolved pulse profiles are shown in Figure \ref{4} (NICER),  and Figure \ref{5} (NuSTAR), respectively.  

The relative amplitude of the evolving pulse profile is expressed in terms of the pulse fraction (PF). PF signifies the X-ray emission from the accretion column (i.e., pulsed emission) and other parts of the
accretion flow or neutron star (NS) surface (i.e., unpulsed emission) \citep{w}. It is expressed in the form, $PF=\;\frac{P_{max}-P_{min}}{P_{max}+P_{min}}$, where $P_{max}$ and $P_{min}$ represent maximum and minimum intensities of the emerging pulse profile respectively. According to the four NuSTAR observations, the PF of 4U 1626-67 exhibits an overall increasing pattern with energy, which is typical of X-ray pulsars \citep{37}. The apparent tendency of PF to decrease around 35-40 keV indirectly justifies the presence of CRSF in the X-ray continuum. Several evidences of decrease in the PF around the cyclotron line energy have been reported in the literature \citep{TSY, 39, 37, tsy, lut}. Figure \ref{6} represents the variation of PF with energy. Recently, \cite {Ferrigno} presented an approach to extract spectral information from the energy-resolved light-curves/pulse profiles in accreting X-ray binary systems utilizing a sample well-known X-ray pulsars including 4U 1626-67. In their investigation, they have considered NuSTAR data during the spin-up state of the source \citep{Camero-Arranz, Jain, Sharma}. They employed a phenomenological polynomial model to characterize the dependence of energy-resolved pulsed fraction and looked for characteristics that match spectral signatures of iron emission or cyclotron absorption lines. A double-humped PF continuum of 4U 1626-67 was observed, which could potentially indicate a modification in the detected emission regions or in the combination of distinct spectral components peaking at various energy bands \citep{Ferrigno, Tsygankov}.  A cyclotron line at about 38 keV in 4U 1626-67 has been verified several times using various observatories and spectrum models \citep{Orlandini, Coburn,  Camero-Arranz, Sharma}. It was observed that the local PF decreases in the spectrum at the known cyclotron line energy.

\section{Spectral Analysis}
The X-ray spectra of 4U 1626-67 were fitted in the broad-band (0.7-50) keV energy range using the NPEX model \citep{Mihara}. We ignored the spectra above 50 keV due to background contamination. The NICER and NuSTAR (FPMA \& FPMB) data were grouped using FTGROUPHA by following the optimal binning approach \citep{Kaastra} with a minima of 25 counts per bin. To account for the instrumental uncertainties and non-simultaneity of the observations, NICER and NuSTAR spectra were simultaneously fitted using a CONSTANT model. We have taken care of the relative normalization factors between the two NuSTAR modules by freezing the constant factor corresponding to instrument FPMA as unity without imposing any constraint on FPMB so as to maintain equivalent count rates in the two instruments. A continuum formula of the following form can be used to describe the continuum spectra of X-ray pulsars adequately:

 F(E) $\propto (E^{-\alpha} + E^{\beta})\times exp {(-E/kT)}$ ,
 where $\alpha , \beta$ and T are positive parameters and k is the Boltzmann constant. The NPEX formula combines negative and positive power laws with a single exponential cutoff factor. The NPEX continuum corresponds to a photon number spectrum for an unsaturated thermal Comptonization in a plasma of temperature T \citep{Sunyaev, Meszaros}. It is the conventional power law with a negative slope for low energies. The positive power law term eventually takes over with increasing energy.

\begin{figure} 
\begin{center}
\includegraphics[angle=-90, scale=0.3]{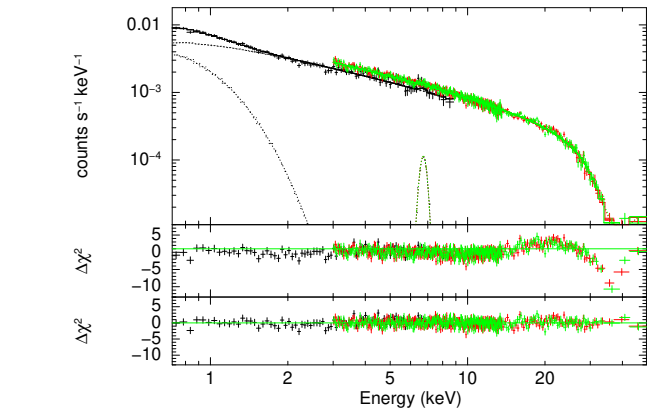}
\end{center}
\caption{The spectra corresponding to joint NICER-NuSTAR observation of 4U 1626-67 in (0.7-50) keV energy range. The top panel is the unfolded spectra, middle panel represents the residuals without incorporating CYCLABS component and the bottom panel represents the residuals after imposing the CYCLABS component for fitting the residual. The spectra is rebinned for representation.}
\label{7}
\end{figure}

The absorption column density is modelled using the TBABS component with abundance from \cite{Wilms}, and the cross-section is taken as \textit{vern} \citep{Verner}. The Galactic survey value of nH \citep{H} is close to that observed with missions having a good soft energy coverage \citep{Camero-Arranz, Beri, D'Ai}. Therefore, we fixed the nH to the galactic value. Since NuSTAR misses the spectra in the soft energy band, we incorporated NICER spectra below 3 keV, and performed simultaneous spectral fitting. The source spectra are associated with soft excess which was modelled using thermal blackbody emission component (BBODYRAD). The black body temperature kT was estimated to be $\sim$ 0.25 keV.  After imposing the continuum model - CONST*TBABS*(NPEX+BBODYRAD), the positive residual accounting for an iron emission line at $\sim$6.8 keV was fitted using a GAUSSIAN component. The negative residuals observed in the X-ray continuum may be interpreted as a cyclotron line. We tried two models (GABS and CYCLABS) dedicated to analyze the geometry of the absorption line as the source is known to exhibit an asymmetric CRSF feature \citep{D'Ai, Iwakirii}. In contrast to the CYCLABS model, which has a pseudo-Lorentzian optical-depth profile, the GABS model has a Gaussian optical-depth profile \citep{Mehara, Makishima}. The Lorentzian line profile, along with its width and depth, represented the spectrum relatively better than the GABS model, despite the fact that both models produced statistically similar fits. Therefore, we employed CYCLABS model to fit the CRSF feature as used in the literature \citep{Sharma}. The centroid energy of the line was estimated to be $\sim$36 keV. The best spectral fit parameters corresponding to the four observations are presented in Table \ref{3}. The flux in the (0.7-50) keV energy range is estimated to be $\sim\;(2.83-4.02) \;\times\;10^{-10} erg\;cm^{-2}\;s^{-1}$, which corresponds to an X-ray luminosity of $\sim\;(3.3-4.9)\;\times\;10^{36} erg\;s^{-1}$ considering a source distance of 10 kpc \citep{Iwakirii}. The joint NICER-NuSTAR spectra corresponding to observation II is presented in Figure \ref{7}.
 
\begin{table*}
\begin{center}
\begin{tabular}{ccccc}
\hline															
Parameters	&	OBS I	&	OBS II	&	OBS III	&	OBS IV	\\
\hline									\\
$C_{NICER}$	&	0.864$\pm$0.002	&	0.854$\pm$0.002	&	 0.830 $\pm$ 0.001	&	0.817$\pm$0.02	\\
$C_{FPMA}$	&	1(fixed)	&	1(fixed)	&	1(fixed)	&	1(fixed)	\\
$C_{FPMB}$	&	0.997$\pm$0.003	&	1.004$\pm$0.005	&	0.998$\pm$0.004	&	1.020$\pm$0.006	\\
$n_{H}\;(cm^{-2})$	&	0.096 (fixed)	&	0.096 (fixed)	&	0.096 (fixed)	&	0.096 (fixed)	\\
kT (BB) (keV)	&	0.24$\pm$0.01	&	0.25$\pm$0.01	&	0.26$\pm$0.01	&	0.25$\pm$ 0.01 	\\
$\Gamma$	&	 0.83$\pm$0.12	&	0.81$\pm$0.14	&	 0.79$\pm$0.12 	&	  0.61 $\pm$   0.10	\\
kT (NPEX) (keV)	&	 7.83$\pm$1.14	&	 7.94$\pm$1.27	&	 7.79$\pm$1.19	&	 7.98$\pm$1.47	\\
Fe line (keV)	&	6.84$\pm$ 0.07	&	6.78$\pm$ 0.07	& 6.79$\pm$ 0.14		&	 6.78$\pm$ 0.28	\\
$\sigma_{Fe}$ (keV)	&	  0.20 (fixed)	&	  0.20 (fixed)	&	  0.20 (fixed)	&	  0.20 (fixed)	\\
$E_{cyc}$ (keV)	&	36.82 $\pm$0.20	&	 36.48$\pm$0.32	&	 36.39 $\pm$0.35	&	36.30 $\pm$ 0.35	\\
$\sigma_{cyc}$ (keV)	&	 7.93 $\pm$ 0.69 	&	7.89 $\pm$ 0.89 	&	7.74 $\pm$ 0.73 	&	 7.62 $\pm$1.32	\\
$Depth_{cyc}$(keV)	&	2.63$\pm$ 0.18	&	  2.46$\pm$ 0.21	&	 2.43$\pm$ 0.26	&	2.45$\pm$ 0.37	\\
Flux ($\times   10^{-10}\;erg\;cm^{-2}\;s^{-1}$)	&	4.02$\pm$0.02	&	3.29$\pm$0.01	&	3.01$\pm$0.02	&	2.83$\pm$0.01	\\
$\chi^{2}_{\nu}$	&	820/778	&	798.42/757	&	784.36/721	&	767.02/745	\\

 \hline
 \end{tabular}
 \caption{The fit parameters of 4U 1626-67 using joint NICER and NuSTAR observations. Flux is estimated within energy range (0.7-50) keV. The absorption column density (nH) is expressed in units of $10^{22} $. The fit statistics $\chi_{\nu}^{2}$ denotes reduced $\chi^{2}$ ($\chi^{2}$ per degrees of freedom). Errors quoted are within 1$\sigma$ confidence interval.} 
 \label{3}
  \end{center}
 \end{table*}

\subsection{Phase-Resolved Spectral Analysis}
\begin{figure}

\includegraphics[angle=0,scale=0.3]{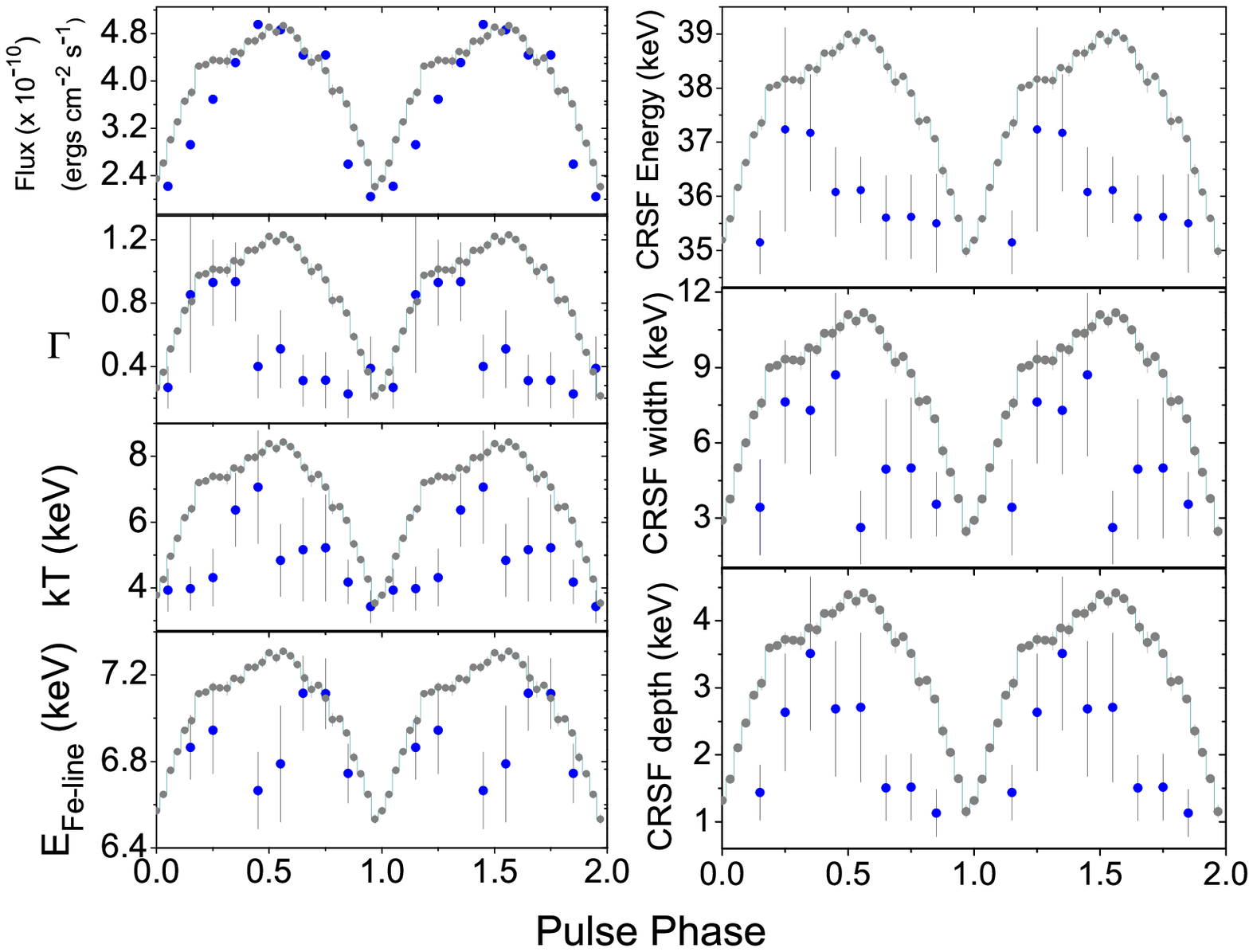}

\caption{Variation of spectral parameters: Flux, Photon Index ($\Gamma$), Plasma temperature (kT), Iron line energy, \& Cyclotron line parameters (energy, width, and depth) with phase for NuSTAR observation I.}
\label{8}
\end{figure}

Phase-resolved spectral study can be efficiently used to compare the spectrum from different portions of the spin cycle and hence, the anisotropic properties of the X-rays emitted by the pulsar around its rotational phase. The pulse profiles of the source reveal an energy dependence. Therefore, it would be intriguing to investigate how the spectral parameters depend on varying the viewing angle of the neutron star. For this purpose, we divided the estimated spin period of the NuSTAR observation (Obs I) into 10 segments and generated the spectra corresponding to each segment. Using the NPEX model, the spectra corresponding to all the phase bins are well fitted. The spectral properties of the source are found to depend upon the pulsed phase. The variation of the spectral parameters relative to the pulsed phase is presented in Figure \ref{8}. The flux has been estimated in the 3-50 keV energy range for each phase bin. The variation of flux with the pulsed phase is comparable to the pulse profile of the source. The flux is found to range in the limit (2.2-4.9)$\times 10^{-10}ergs\; cm^{-2}\; s^{-1}$. The power-law photon inCyclotron lines have been detected in several X-ray pulsars. The physical characteristics of the cyclotron lines and any novel phenomena that may be associated to it are the current areas of interest for X-ray astronomers. This includes the long-term evolution or luminosity dependency of the cyclotron line energy.dex is found to follow an increasing trend along the pulse profile during the initial phase bins and is found to drop significantly in the later phase bins. It ranges in the limit (0.22-0.93) with a maximum along the phase interval (0.3-0.4), indicating a softer spectrum at that phase bin. It attains a minima close to the minima of the pulse profile in the phase interval (0.8-0.9). The plasma temperature (kT) also varies with the pulse phase, and ranges from 3.42 to 7.06 keV. It is found to be maximum near the peak emission of the pulse profile and minimum at the end phase bin coincident with the minima of the pulse profile. The iron emission line is prominently observed along seven phase bins. The centroid energy of the Fe line exhibits a complex evolution with the pulsed phase and differs from the phase-averaged value by about 4 $\%$. The centroid energy of the Fe line marks a peak value of $\sim$ 7.12 keV along the phase interval (0.6-0.7) and a lowest value of $\sim$ 6.66 keV along the phase interval (0.4-0.5).

The evolution of the CRSF parameters relative to pulsed phase is observed. The CRSF energy exhibits variations with respect to the pulsed-phase due to sampling different heights of the line forming region. The centroid energy of cyclotron line is found to deviate by around (3-4) $\%$ relative to the phase-averaged value. The maxima and minima of the cyclotron line energy are observed along the phase intervals (0.2-0.3), and (0.1-0.2) respectively. Except for the phase intervals (0.2-0.3), and (0.3-0.4) where a noticeable increase in the line energy is evident, it remains generally steady over the initial and final phase intervals. We notice a phase shift of 0.2 in the maxima of the cyclotron line energy to that of the pulse profile. The CRSF width and depth exhibit complex fluctuation with the pulse phase. The width and depth of the cyclotron line range in the limit (1.58-3.72) keV and (2.23-9.15) keV, respectively. The cyclotron line is not detected during some rotational phases which is evident from Figure \ref{8}. 

\section{Discussion \& Conclusion}
The spectral and temporal features of the persistently active LMXB source 4U 1626-67 are studied using recently observed NuSTAR and NICER data following its recent torque reversal to the spin-down phase. Coherent X-ray pulsations are detected at $\sim$ 7.67 seconds. The pulse profile has changed significantly from what was found before the recent torque reversal, which may be associated with a change in the geometry of the accretion column. The spin-up and spin-down profiles of the source were recently examined by \cite{Sharma}. The double-peaked structure of the pulse profile during the previous spin-up episode of the source is not observed during the current spin-down episode. The pulse profiles are consistent with the previously established torque-state dependencies \citep{Beri}.  For spin-up phases, the absorbing structures are narrower than the emission beam, resulting in two peaks, while during the spin-down phase, the accretion stream absorbs most of the hotspot emission. However, the strength of the emission from the hotspot and the strength of the accretion stream's absorption differ, with the absorption being relatively stronger during the spin-down phase. This high-energy component does not vary significantly during the spin-up or spin-down episodes. 4U 1626-67 goes through various phases, each of which requires a different size, shape, and geometry of the polar hotspot regions and accretion stream. The accretion process from the inner disc to the surface of the neutron star takes place in various modes that are steady over a long period of time. The X-ray spectra and varying pulse patterns of 4U 1626-67 indicate transitions between such stable accretion states with varying accretion flow geometries. According to \cite{van Kerkwijk}, a radiation-pressure-induced warping of the inner accretion disc may turn retrograde and produce a negative accretion torque of a comparable magnitude. The energy-dependent pulse patterns can possibly be caused by a different accretion mode during the spin-down phase compared to the spin-up phase. The spin-up state is associated with a stable accretion column that feeds the neutron star at a uniform accretion rate. However, the accretion during the spin-down phase is most likely from a more clumpy material, which justifies the existence of QPOs during this phase, presuming QPOs to be a manifestation of clumps in the inner accretion disk. The pulse profiles of 4U 1626-67 are  known to be strongly energy-dependent, differing in morphology during the spin-up and spin-down states. The complex morphology of the pulse profiles may be explained by a hybrid fan beam and pencil beam emission pattern. At energies lower than $\sim$ 10 keV, both beam patterns contribute at similar levels to the pulse profile, while at higher energies, the pencil beam emission pattern is more dominant \citep{Iwakiri, Marshall}.

The PDS of the source reveals the presence of a broad $\sim$ 46.8 mHz QPO in all the four NuSTAR observations which is in accordance with the recent reports of \cite{Sharma}. This feature only appears during the spin-down episodes and is not present during the spin-up phase of the source \citep{Kaur, Jain, Beri}.  Considering quasi-periodic oscillation frequency is due to the beating between the pulse frequency and the Keplerian motion at the magnetospheric radius, the magnetic field strength of the source may be approximated by the relation $B_{12} \sim 5.5{\sqrt{L_{37}}}$ \citep {Shinoda}, where $L_{37}$ is the X-ray luminosity in units of $10^{37} ergs\; s^{-1}$. Assuming a source distance of 10 kpc, it corresponds to a magnetic field strength of $\sim$ 3.3 $\times10^{12}$ G, consistent with our measurements considering the cyclotron line energy.

Accretion torque is associated with the magnetic coupling between the neutron star and the accretion disc around it. Changes in the accretion torque may be related to the source luminosity and spin transition as it is associated with the mass accretion rate \citep{Ghosh}. Long-term cycles in the supply of matter from the companion star may be the reason behind the time scale for torque reversals in 4U 1626-67. The current torque reversal is characterized by a significant decrease in flux, and the source is spinning down at the rate 0.00045(4) s\;$yr^{-1}$. The torque reversal encountered by the source suggests that it may be spinning close to its equilibrium period , where the spin frequency matches the Kepler frequency at the magnetosphere. A neutron star spinning with a shorter period than the equilibrium period cannot accrete easily and encounters a strong spin-down torque known as propeller effect \citep{Illarionov}. The angular momentum transfer in 4U 1626-67 is relatively smooth in contrast to most other accreting pulsars \citep{Bildsten}. Low-mass X-ray binary pulsars and persistent X-ray pulsars with supergiant companions do not exhibit a strong correlation between the torque on the neutron star and the X-ray luminosity or mass accretion rate \citep{Bildsten}. The standard model of accretion onto the magnetized neutron star is, therefore, not fully applicable in such sources. Most persistent pulsars do not exhibit a distinct correlation between the mass accretion rate and accretion torque. In the case of 4U 1626-67, sharp torque reversals have been identified, each of which is connected to a substantial spectral transition. A thorough analysis of the long-term X-ray flux and spin evolution history indicates that the transfer of torque in 4U 1626-67 may also be influenced by other factors in addition to the mass accretion rate. The average mass accretion rate ($\dot{M}_{avg}$) may be approximated using the relation:

$P_{spin} \; \sim \;[\frac{10^{-10} \textup{M}_\odot}{\dot{M}_{avg}}]^{3/7} [\frac{\mu}{10^{30}}]^{6/7}$, 
where the symbols have their conventional meanings. For the estimated magnetic field strength considering the cyclotron line energy, $\dot{M}_{avg}$ is found to be $\sim 1.1 \times 10^{-9}\textup{M}_\odot$. We confirm the estimated mass accretion rate during the present observations to be $\sim3\times 10^{-10}\textup{M}_\odot$ \citep{Sharma} which is lower than $\dot{M}_{avg}$, suggesting a spin-down torque.

 The PF, which represents the fraction of photons contributing to the modulated part of the flux, reveals interesting characteristics associated with the source. The variation of PF with energy is consistent with the presence of characteristic absorption features observed in the X-ray continuum of the source. The PF follows an overall increasing trend with energy. The increasing trend of PF with energy was explained using a toy model \citep{landt} which suggests that the X-ray emitting region becomes more compact with increasing energy, enhancing the pulsed emission of the neutron star. It is characterized by a drop around specific energies which accounts for the prominent CRSF evident in the source spectra. The non-monotonic dependence of the PF on energy about the cyclotron line energy has been reportedly observed in several X-ray pulsars \citep{38,37,39}. An approach to extract spectral information from energy-resolved light-curves/pulse profiles in accreting X-ray binary systems was recently reported by \citep{Ferrigno}. The approach used a sample of well-known X-ray pulsars, including 4U 1626-67 accreting in the spin-up state. A double-humped PF continuum of 4U 1626-67 was observed indicating that the identified emission zones have changed, or different spectral components that peak at different energy bands have been combined \citep{Ferrigno, Tsygankov}.

 Since NuStar's low energy threshold starts at $\sim$ 3 keV, we incorporated concurrent NICER spectra below 3 keV, and performed simultaneous broad-band spectral fitting. The joint NICER-NuSTAR broadband spectra of the source has been well fitted by using a single continuum model and the corresponding fit parameters are presented in Table \ref{3}. We employed the generally accepted spectral model for 4U 1626-67, revealing a blackbody component (kT $\sim$0.25 keV) in the current torque state,  which was not detectable with NuSTAR in the recent study of the source \citep{Sharma}. Considering kT = 0.25 keV, the blackbody luminosity is estimated to be $\sim 6.41 \times 10^{35} erg\; s^{-1}$ , which is consistent with that reported by \cite{Sharmaa}. The current spin transition of the source from spin-up (in 2015) to spin-down (in 2023) is accompanied by significant changes of spectral parameters. The black body temperature during the current phase has decreased from $\sim$ 0.5 keV to $\sim$ 0.25 keV, indicating an increase in the effective area of the black-body source during the low luminosity phase. Variations of the effective area of the black-body source may be due to the restructuring of the accretion channel. The channel parameters depend on the geometry and physical conditions in the accretion flow at the magnetospheric boundary and the mode by which the material enters the neutron star's magnetic field. In addition to an iron emission line at $\sim 6.8$ keV, an absorption feature is prominently observed in the X-ray continuum.  The average flux of the source in (0.7-50) keV energy range was found to be $\sim3.3 \;\times\;10^{-10}\;erg\;cm^{-2}\;s^{-1}$, which represents a luminosity of $\sim 3.9 \;\times\;10^{36}\;erg\;s^{-1}$, assuming a distance of 10 kpc. It implies that 4U 1626-67 lies in the intermediate luminosity range of accreting X-ray pulsars that may be characterized by a hybrid accretion geometry \citep{Becker}. We estimate a magnetic field strength of $\sim$ 3.10 $\times 10^{12}$ G for the cyclotron energy of 36 keV.The observations are characterized by decreasing flux/luminosity. Generally, the cyclotron line energy in most sources is dependent on the accretion state. The cyclotron line parameters are known to vary with luminosity \citep{Staubert}. However, we observe no significant variation in the centroid energy of the cyclotron line over the period of the observations suggesting that the cyclotron
line energy in 4U-1626-67 is independent of the accretion state. Cyclotron lines have been detected in several X-ray pulsars. The physical characteristics of the cyclotron lines and any novel phenomena that may be associated with them are the current areas of research interest in X-ray astronomy. It includes the long-term evolution and luminosity dependence of the cyclotron line energy. Regarding the present understanding of the cyclotron line parameters in accreting X-ray pulsars, refer to \cite{Staubertt}. 

The geometry of the emission regions near the neutron star and the structure of its magnetic field may be analyzed using pulse-phase-resolved spectroscopy. The spectral parameters of the source are found to exhibit significant variabilty relative to the pulsed phase. The cyclotron line is not observed along some rotational phases. The non-detection of the cyclotron line in some phases might be due to the limited statistics as those phases lie at the bottom of the pulsed emission.

\section*{Acknowledgements}
This research work is carried out using publicly available data provided by NASA HEASARC data archive. The NuSTAR data is provided by NASA High Energy Astrophysics Science Archive Research Center (HEASARC), Goddard Space Flight Center. Additionally, data from Fermi-GBM pulsar monitor are utilised. We acknowledge IUCAA Centre for Astronomy Research and Development (ICARD), Department of Physics, NBU, for extended research facilities. The authors would like to express their gratitude to the anonymous reviewer for providing insightful comments that improved the content of the manuscript.

\section*{Data availability}
 
The observational data used for performing this research work can be accessed from the HEASARC data archive that is publicly available.









\bsp	
\label{lastpage}
\end{document}